\documentclass[fleqn,usenatbib]{mnras}
\usepackage{newtxtext,newtxmath}
\usepackage[T1]{fontenc}

\DeclareRobustCommand{\VAN}[3]{#2}
\let\VANthebibliography\thebibliography
\def\thebibliography{\DeclareRobustCommand{\VAN}[3]{##3}\VANthebibliography}

\usepackage{graphicx}	
\usepackage{amsmath}	
\usepackage{threeparttable}
\usepackage{ulem}

\title{The disrupting and growing open cluster spiral arm patterns of the Milky Way}

\author[Liu et al.]{
Xiaochen Liu,$^{1}$
and Zhihong He$^{1}$\thanks{E-mail: hezh@mail.ustc.edu.cn}
Yangping Luo$^{1}$
and Kun Wang$^{1}$
\\
$^{1}$School of Physics and Astronomy, China West Normal University, No. 1 Shida Road, Nanchong 637002, People's Republic of China
}

\date{Accepted XXX. Received YYY; in original form ZZZ}

\pubyear{2024}

\begin{document}
\label{firstpage}
\pagerange{\pageref{firstpage}--\pageref{lastpage}}
\maketitle

\begin{abstract}
Star clusters provide unique advantages for investigating Galactic spiral arms, particularly due to their precise ages, positions, and kinematic properties, which are further enhanced by ongoing updates from the astrometric data. In this study, we employ the latest extensive catalogue of open clusters from Gaia DR3 to examine the positional deviations of clusters belonging to different age groups. Additionally, we employ dynamical simulations to probe the evolutionary behavior of spiral arm positions. 
Our analysis reveals an absence of a theoretical age pattern in the spiral arms traced by open clusters, and the pattern speeds of the spiral arms are consistent with the rotation curve. Both of these results do not align with the predictions of quasi-stationary density wave theory, suggesting a more dynamic or transient arm scenario for the Milky Way. From this perspective, combined with vertex deviation estimates, it appears that the Local arm is in a state of growth. In contrast, the Sagittarius-Carina arm and the Perseus arm exhibit opposing trends.
Consequently, we speculate that the Galactic stellar disk does not exhibit a grand-design spiral pattern with a fixed pattern speed, but rather manifests as a multi-armed structure with arms that continuously emerge and dissipate.
\end{abstract}





\section{Introduction}

Spiral arms are the typical morphological features of spiral galaxies and serve as the primary sites for star formation within these galaxies. However, the formation and evolution of spiral arms have been a topic of much debate ~\citep{Dobbs14}. Two widely discussed theoretical models for spiral arms are the quasi-stationary density wave theory~\citep{Lin64,Lin66} and the dynamical spiral arm formation theory~\citep{Sellwood84}. One of the fundamental differences between these two models lies in whether the spiral arm patterns are dynamically changing. 
The former predicts that galaxies have grand-design density wave spiral patterns that remain steady, causing gas and stars of different ages to have position gradients on the arms ~\citep{Roberts69,Shu16}. The latter suggests that the spiral arm pattern is transient, consisting of flocculent or multi-arm features that continuously disappear and recurrence~\citep[e.g.][]{Sellwood14,Sellwood19b}. 
However, observational and simulation evidence supports (or contradicts) both models, leaving the question of whether spiral arm patterns are stable or not unresolved~\citep{Sellwood11}.

As the home galaxy of humanity, the existence of spiral arm structures in the Milky Way has been well-known since the last mid-century~\citep{Morgan53}. Radio observations have revealed magnificent HI spiral arm structures that extend beyond 25 kpc from the Galactic disk, overcoming the effects of interstellar extinction~\citep{Levine06}. Within the Milky Way, the active star formation illuminates the profile of the spiral arms, visible through the distribution of HII regions~\citep{Georgelin76}, CO molecular gas~\citep{Cohen80,Dame11,Sun15}, and young stellar objects such as classical Cepheids~\citep{Skowron19}, OB stars~\citep{Xu18,Cheng19}, or comoving groups~\citep{Kounkel20}. With advancements in spectroscopic surveys and astrometric measurements, the details of the Milky Way's spiral arms, particularly those near the solar vicinity, are being progressively revealed~\citep[e.g.][]{Hou14,Xu16,Reid19,Collaboration23}. This provides better opportunities to investigate the details of spiral arms, particularly in terms of verifying theoretical predictions.

However, many arm tracers face limitations. For example, interstellar medium observations are affected by distance uncertainties, VLBA observations have limited samples, and individual stellar astrometrics have larger uncertainties than star clusters. In contrast, open clusters' astrometric measurements can be derived from the statistical properties of member stars, resulting in higher distance and kinematic precisions than for individual objects. Moreover, color-magnitude diagrams of clusters can effectively distinguish different age groups of clusters. With continuous releases of Gaia astrometric data over the past five years~\citep{Collaboration18,Collaboration21}, the precision of stellar parallaxes and proper motions has greatly improved, and the parameters of cluster member stars are becoming increasingly accurate. 

Therefore, open clusters (hereafter OCs) with wider age ranges have become convenient targets for observing structure and the evolution of spiral arms in the Galaxy~\citep{Dobbs10b}. In the Gaia DR2 era, based on re-identified and newly found OCs~\citep[e.g.][]{Gaudin18,Sim19,Liu19,Ginard20,He21a,Hunt21}, the structure and kinematic features of the Milky Way's spiral arms were revealed~\citep{Gaudin20a}. In our previous work~\citet{He21b}, we have investigated the age pattern of the Milky Way's spiral arms using OCs and found possible deviations in the positions of different age ranges within the Local arm and Sag-Car arm. However, due to the limited sample size, this trend was not significant, and the traced arm was relatively short, and no obvious age pattern found in Perseus arm. Another previous work~\citet{Ginard21} (hereafter CG21) has suggested that age patterns (also known as age gradients) do not exist in the spiral arms traced by OCs.

With the discovery of new OCs in Gaia DR3~\citep[e.g.][]{Ginard22,Hunt23,He23b}, the range of the Galactic disk traced by OCs extends to 15~kpc from the Galactic center~\citep{He23a}, and more line-of-sight velocity data become available~\citep{gaia23}, presenting new opportunities for the study of spiral arms in the Milky Way. Therefore, in this study, our aim is to comprehensively investigate the Milky Way's spiral arms using the new OC sample, with a particular focus on the positional distribution differences among clusters of different age groups. By applying dynamic models, we aim to discuss whether the spiral arms are steady or transient. The findings of this study will contribute to our understanding of the formation and evolution of spiral arms in the Milky Way and have implications for the broader study of spiral galaxies.

Next, in Section~\ref{sec2}, we introduce the tracers used in our study. In Section~\ref{sec3}, we present the positional differences among clusters of different age groups and discuss these results in the context of other studies. We then discussed the pattern speed (Section~\ref{sec4}) and vertex deviation (Section~\ref{sec5}) of each spiral arm. Finally, we provide a conclusion in Section~\ref{sec6}.


\section{Data}
\label{sec2} 
In our previous research, we used an OC catalogue compiled by ~\cite{Gaudin20a} (hereafter CG20), based on the Gaia DR2 dataset~\citep{Collaboration18}. This catalogue encompassing 1,827 OCs with each containing over 20 member stars, and constituted a fundamental resource for our analysis of the spatial distribution and dynamics of these stellar congregations within the Milky Way. Despite its utility, the coverage provided by this sample regarding the Galactic spiral arms was relatively limited, with younger star clusters particularly underrepresented: accounting for only 693 OCs younger than 100 Myr, that may introduces biases in the statistical analysis of the sample due to the incomplete mapping of such clusters. The advent of the Gaia EDR3~\citep{Collaboration21} offers improved accuracy of astrometric measurements, including parallaxes, proper motions, and photometric data across over 1.5 billion stars, thereby significantly enhancing the capability to identify new OCs. Furthermore, the subsequent Gaia DR3~\citep{Collaboration23} expanded the dataset further by including additional radial velocity measurements. This influx of new data has dramatically expanded the pool of identified OCs, enriching the sample with younger clusters that were previously uncharted, and promising to fill the previously noted gaps in our statistical analyses. This enhanced dataset not only provides a more comprehensive overview of the Milky Way's structure but also sets the stage for a refined understanding of the dynamics and evolutionary history of its stellar populations.

Subsequent work expanded upon the initial CG20 dataset by integrating additional discoveries from the Gaia DR2/EDR3 era, and incorporating approximately 3,500 newly identified star clusters~\citep{Ginard22,He21b,He2022a,He2023a,He23b}, all maintaining a minimum of 20 members per cluster. This updated catalogue now boasts 5,340 reliable OCs, substantially augmenting the CG20 collection by threefold. We then cross-matched the clusters in this catalogue, removing 165 duplicate clusters and clusters without age information. In addition, we have added 279 neighboring star clusters that were not previously included~\citep{He22b}. Moreover, our analysis was expanded by adding 412 clusters identified as reliable OCs, based on a comprehensive cross-match with the OC catalog presented by~\citet{Hunt23}. This enhancement significantly broadens the scope for detailed studies of the Milky Way's spiral arm structures, particularly those in close proximity to our solar system. 

In this analysis, we categorize the 5,866 OCs into different age groups: 1,144 clusters younger than 20~Myr, 633 clusters aged between 20~Myr and 50~Myr, 915 clusters aged between 50~Myr and 100~Myr, and 3,174 clusters that are older than 100~Myr. Our focus primarily lies on star clusters younger than 100~Myr, which, numbering 2,692, marks a substantial increase in sample size: over threefold compared to the dataset analyzed by~\citet{He21b} and CG21. 
Based on Gaia DR3 data~\citep{Collaboration23}, 4,038 OCs from this sample have radial velocity measurements with uncertainties of less than 5 km~s$^{-1}$, all the clusters have parallax-based distance.
This significant expansion of the dataset is invaluable for discerning age patterns within the spiral arms, as well as for retracing and understanding the evolutionary positional shifts of the star clusters.
In this study, the derivation of cluster ages primarily stems from isochrone fitting applied to the color-magnitude diagram, utilizing the PARSEC isochrones~\citep{Bressan12}. Approximately one third of our sample builds upon ages from CG20, while our previous work contributes almost two thirds. We observe a systematic deviation of 0.1~dex (logarithmic age) between ages derived from CG20's methodology and our fitting approach~\citep{He22b}. To address potential impacts of age uncertainties on the results, we have introduced a random error of $\pm$~0.1~dex to the ages of all clusters considered in subsequent analyses.

\section{OC spiral arm age patterns}
\label{sec3}

\begin{figure*}
\begin{center}
	\includegraphics[width=1.59\columnwidth]{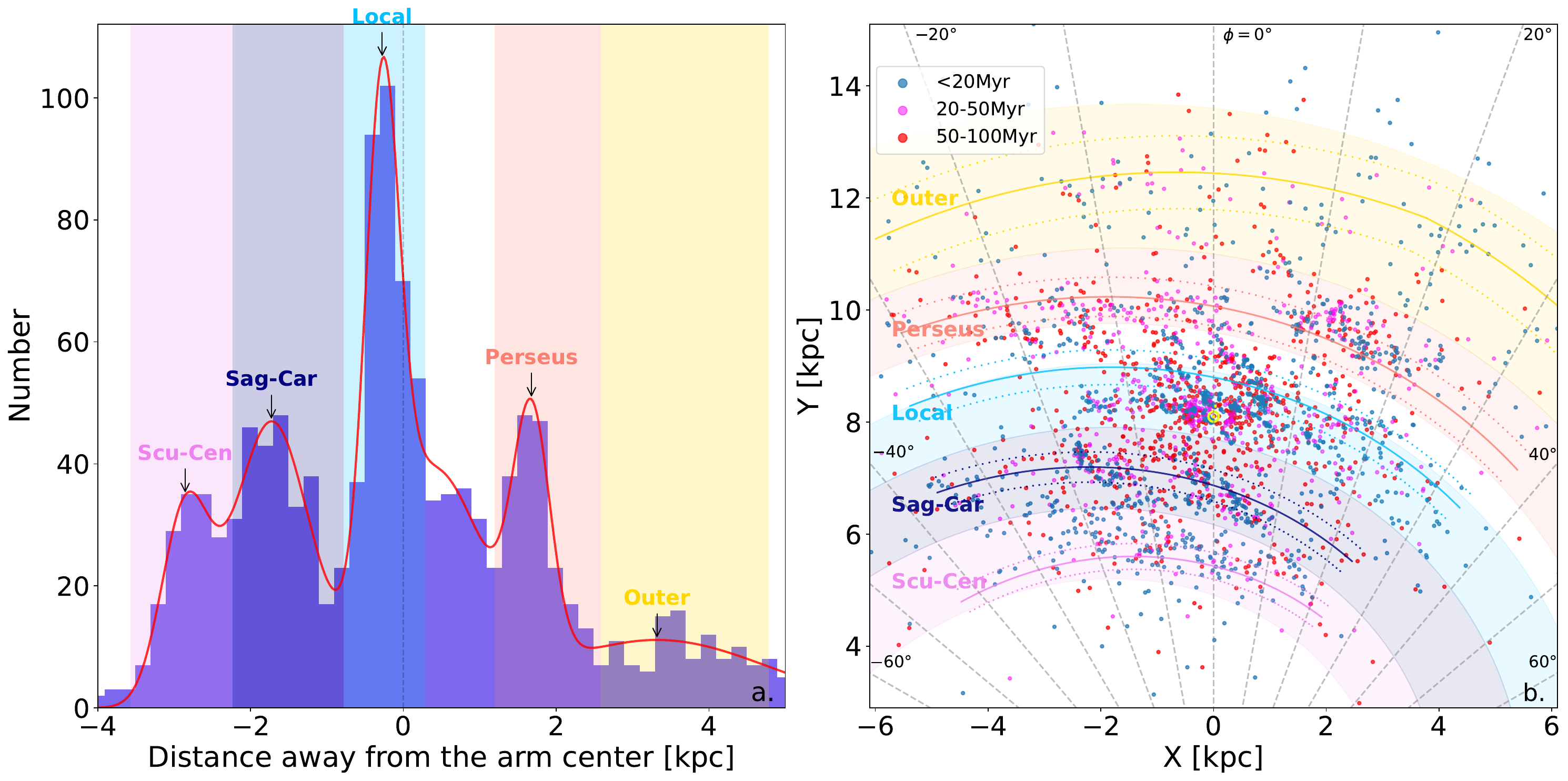}
    \caption{Left panel: Distribution of the minimum distance of OCs younger than 30~Myr from the Local arm~\citep{Reid19}, with the Y-axis representing the cluster numbers fitted with the Gaussian curves. The background highlights the range of the spiral arms we considered. Right panel: The projection positions of OCs of different age ranges on the Galactic disk are differentiated by various colors, displaying the areas of the spiral arms in the same colors as in the left panel. The sun is positioned at (0, 8.15)~kpc, and the colored curves represent different spiral arms following~\citet{Reid19}. Azimuthal angle at $\phi$ = 0$^{\circ}$ denotes the line from the Galactic center towards the sun, adopting a clockwise positive direction.}
    \label{fig_0}
\end{center}
\end{figure*}

Previous N-body simulations have indicated that young stars with ages up to 30~Myr can still exhibit the spiral arm pattern and with no significant offset with gas spirals~\citep{Wada11}, and studies have demonstrated that even older OCs with ages between 50-100 Myr can effectively serve as tracers for the spiral arm structure~\citep[e.g.][]{Gaudin18,Ginard21}. In this study, we utilize OCs with ages below 100~Myr to trace the positions of the spiral arms, dividing them into three age groups: <20~Myr, 20-50~Myr, and 50-100 Myr. 
We adopted the spiral arm model proposed by \citet{Reid19} to classify OCs associated with various spiral arms. 
Initially, we calculated the minimum distance from each cluster to the Local Arm as defined in \citet{Reid19}, with the zero distance representing the position of the Local Arm.
Through fitting the histogram contours with a multi-Gaussian mixture model, we then established the boundaries of five spiral arms.  
In the majority of spiral arms, the defined region extends 2-sigma outward from the center of the corresponding Gaussian curve. However, in specific areas where the 2-sigma range fails to adequately encompass all relevant regions or overlaps with adjacent spiral arms, we adjust the range to 1-sigma or 3-sigma as necessary.
As depicted in the left panel of Figure~\ref{fig_0}, the background highlights the sigma ranges considered for each spiral arm, with minor discontinuities or overlapping regions omitted for clarity: [-3, 2] for the Scutum-Centaurus (Scu-Cen) arm, [-1, 1.5] for the Sagittarius-Carina (Sag-Car) arm, [-3, 2] for the Local arm, [-2, 3] for the Perseus arm, and [-0.5, 1] for the Outer arm. Insignificant variations (less than 0.5 sigma) in the sample interval have a negligible impact on the subsequent calculation results. 
It should be noted that the inter-arm region between the Local arm and the Perseus arm exhibits a "spur"-like structure, which we have excluded from both the Local arm and the Perseus arm when calculating the pattern speed in Section~\ref{sec4}.
While the Outer arm and Scu-Cen arm possess fewer samples, our primary focus lies on the Perseus arm, Local arm, and Sag-Car arm (right panel in Figure~\ref{fig_0}).

\begin{figure*}
\begin{center}
	\includegraphics[width=1.59\columnwidth]{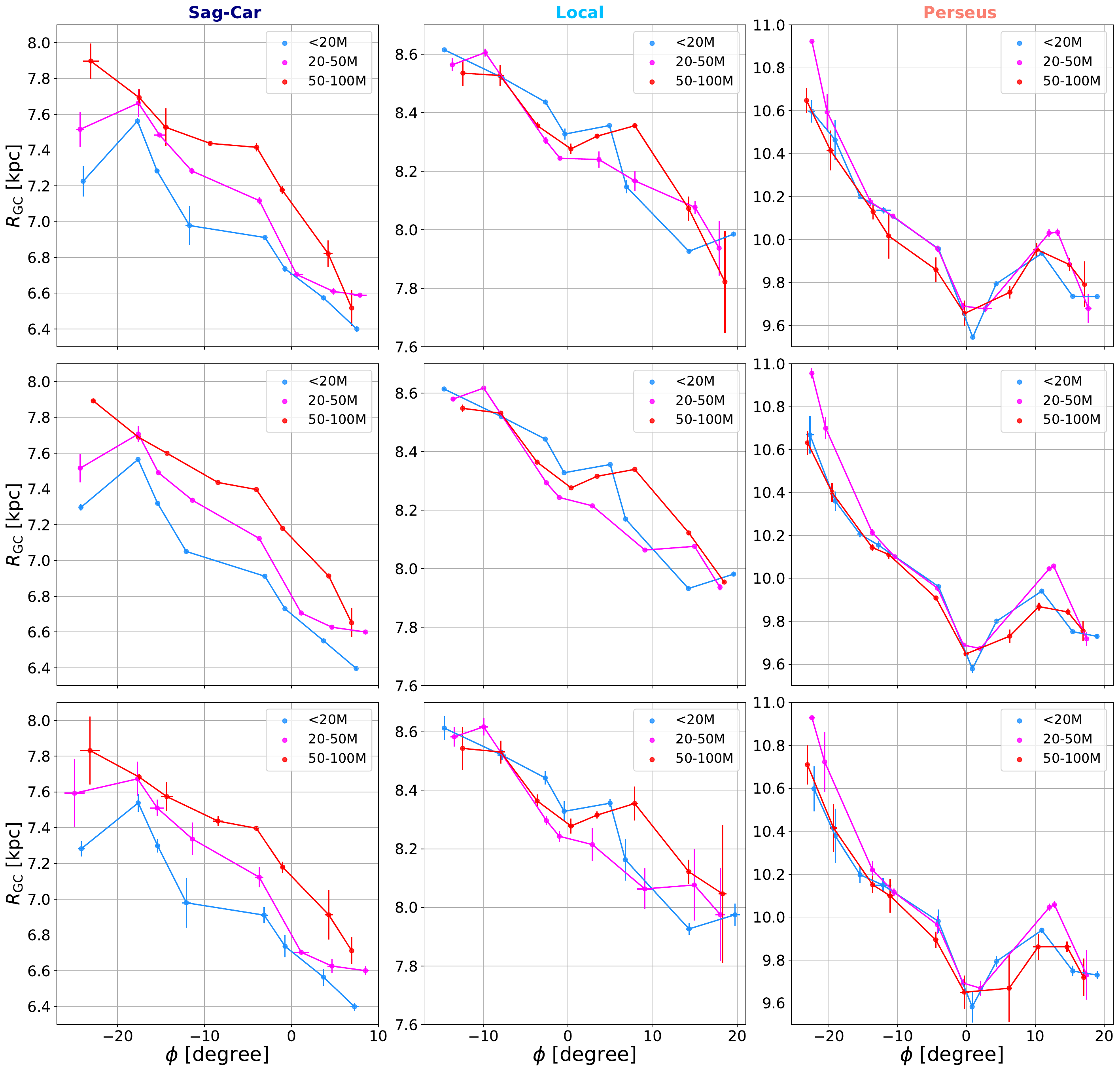}
    \caption{Distribution of azimuthal angles of OCs by age within designated spiral arm regions, plotted against Galactocentric distance R$_{GC}$, with cluster counts in bins $\geq$~10 OCs. Upper panels: Within the age uncertainty range of each star cluster, we randomly select a value to serve as the new age parameter for the cluster. We calculate the results based on this sample and repeat the process 1000 times, ultimately deriving the median and dispersion for each bin. Middle panels: Similar to the upper panels, but the distinction lies in drawing random values according to different distance uncertainties instead of age uncertainties. Lower panels: Within each bin, we apply the bootstrap method, selecting 70$\%$ of the star clusters with replacement. This cycle is repeated 1000 times. We then compute the median and dispersions of these 1000 outcomes to establish our final results and associated errors.}
    \label{fig_1}
\end{center}
\end{figure*}

In previous studies, due to the limited sample size, we analyzed all OCs associated with each individual arm when investigating their positions~\citep{He21a}. After improvement, in this study, we examine the positions along the azimuthal angle $\phi$ (with bins of 10 degrees) with respect to the Galactic center. We omit bins containing fewer than 10 clusters from our analysis. The results of the positional differences for the various arms are displayed in three panels of Figure~\ref{fig_1}. Notably, clusters situated on the Sag-Car arm show a significant and systematic spatial offset between young and old clusters, ranging from 0.1 to 0.5~kpc. 
The youngest (<~20 Myr)  group and the oldest (50 - 100 Myr) portion of the Local arm exhibit a weak offset for radii less than 8.3~kpc (and $\phi$ = 5$^{\circ}$ - 20$^{\circ}$ ). 
However, beyond this radius, the weak offset may disappear and/or undergo a shift in direction, and the middle-aged group (20 - 50 Myr) is not positioned between the youngest and oldest groups.
Consequently, identifying a specific age pattern within the Local arm is challenging.
These findings align with the positional deviations observed in~\citet{He21a} for different tracers along the spiral arms and show enhanced distinctiveness.

Additionally, as depicted in the sketch map presented in Figure~\ref{fig_2}, the magnitudes of the offset for the peak positions of the OCs on the Sag-Car arm align with the predictions of the quasi-stationary density wave theory. According to this theory, when the spiral arm is located near the CR, the offset magnitude is not pronounced relative to the width of the arm. However, as the azimuthal angle deviates further from the CR, the offset becomes more evident. 
Furthermore, within the CR, the older portion (red curve) of the spiral arm distribution exhibits on the far side from the Galactic center, while the trend is reversed outside the corotation radius. 
In quasi-stationary density wave theory, the direction of the age gradient in the age pattern within spiral arms changes at the CR~\citep{Shu16}. Based on this theory, if the age offset observed along the Sag-Car arm but not in the Local arm is due to differences between the spiral arm pattern speed and the cluster rotation speed, then the CR should lie between these two arm regions.

In previously published literature, it is noted that the CR of the Milky Way's spiral arm pattern and the Galactic disk occurs near the sun~\citep[e.g.][]{Amores09,Michtchenko18,Dias19}. Previous research by~\citet{Hou15} explored the offsets of gas and stellar components in the inner Galactic disk, revealing clear angular deviations for tracers on the Scu-Cen arm and indicating similar offset indications on the Sag-Car arm. Subsequently,~\citet{Veselova20} also identified young and old Cephieds located in different positions in nearby spiral arms. These findings are not inconsistent with our observations on the Sag-Car arm, which indicate a positional offset between young, middle-aged, and old OC groups.
Thus, based solely on these morphological observations, the offsets seem to corroborate the presence of age patterns in Sag-Car arm.

Nevertheless, according to predictions stemming from the steady arm view~\citep{Shu16}, the Perseus arm should exhibit a more pronounced age pattern on a larger scale, which is not observed in our analysis. As shown in right panels in Figure~\ref{fig_1}, there are no significant or systematic positional differences among the young and intermediate-aged OCs on the Perseus arm. Similar contradictions were also found in the offset of red clump stars and young objects~\citep{Lin22}. Furthermore, the offset of the age pattern on the Sag-Car arm does not demonstrate a clear trend of increasing magnitude with increasing distance from the potential CR. 
Moreover, within the Local arm, the age pattern does not exhibit a progression from older to middle-aged, and then to younger age.
These characteristics pose a challenge to the interpretation of the Milky Way spiral arms within the framework of the quasi-stationary density wave theory. 
From a morphological perspective, within 5~kpc from the Sun, particularly in the inner disk, the Milky Way exhibits multiple fragmented segments of spiral arms, potentially indicating a multi-armed configuration in our Galaxy~\citep{Xu23} rather than the classical two-armed grand design. Hence, it is not entirely convincing to solely consider the Sag-Car arm as a major arm generated by a density wave, while excluding the Local arm and the Perseus arm.

This situation bears similarities to recent observations in external galaxies, where age patterns are found closer to the inner galactic regions, but are inconsistent in the outer arms~\citep{Garner24}; and age patterns could present or absent in different galaxies~\citep{Choi15,Shabani18,YuYu18,Peterken19}. Consequently, there is an increasing tendency to believe that the density wave alone cannot explain the differences in the spiral arm patterns exhibited by these galaxies. When considering these cumulative characteristics, we find no clear evidence of the theoretically predicted age pattern based on the distribution of OCs in Milky Way, which is support to conclusion of CG21.


\section{Spiral Arm pattern speed}
\label{sec4}

To further investigate the aforementioned discussions, we simulated the orbital evolution of these OCs, using a $Python$ package and the Galactic gravitational potential model MWPotential2014 in
$Galpy$~\citep{Bovy2013,Bovy2015} that specifically designed for Galactic dynamics calculations. 
Additionally, we utilized the DehnenBarPotential gravitational potential model~\citep{Dehnen99,Monari16} to examine the effect of the Galactic bar on the spiral arm pattern speed.

Recent studies by CG21 make use of OCs to derive pattern speeds of various spiral arms in the Galactic disk. They observed that these pattern speeds generally correlate with the Galactic rotation curve as the Galactocentric distance increases; however, these findings diverge from those reported by~\citet{Dias19}. To further investigate, we have expanded upon this research by employing a larger and more comprehensive sample of star clusters to calculate the pattern speeds across different spiral arms.
We utilized the computational approach for determining pattern speeds as outlined by~\citet{Dias05}, which was also referenced extensively in the studies by CG21 and~\citet{Joshi23} (JM23). The methodology involves several key steps for each spiral arm:
\begin{itemize}
\item[\textbullet] Assuming that OCs are born within spiral arms — implying that the birth position of each OC indicates the location of a spiral arm at the time of the OC's formation — we retrace the birthplaces of star clusters in a specified segment of the arm, typically spanning less than 50-80~Myr.
\item[\textbullet] Assuming that the pattern of the spiral arm remains unchanged, we apply a range of pattern speeds to evolve these birthplace coordinates to their expected present-day position.
\item[\textbullet] The present-day positions of the clusters are then compared across various pattern speeds with the contemporary spiral arm configuration. The optimal pattern speed is determined by the degree of alignment between the current positions of the clusters the expected present day arm pattern.
\end{itemize}

In this analysis, we utlized clusters within three age ranges: (10, 50)~Myr, (50, 80)~Myr, and (10, 80)~Myr. We carried out 1000 iterations of above analytical process, taking into account the variability due to random sample selection, and the uncertainties in age, position, radial velocity, and proper motion. The computed mean and standard deviation from these iterations were used to determine the final pattern speed values and their uncertainties for each spiral arm. 
The results are presented in Table~1, demonstrate that variations in the sample have the most significant impact on the pattern speed, with errors mostly ranging from 0.5 to 2.5~km~s$^{-1}$ across different spiral arms. The next error source is the cluster age, with the pattern speed error due to age being about one-third to one-half of that from the sample variations. In contrast, the errors arising from distance, radial velocity and proper motions are minimal and can be largely disregarded.

\begin{figure*}
\begin{center}
	\includegraphics[width=0.95\columnwidth]{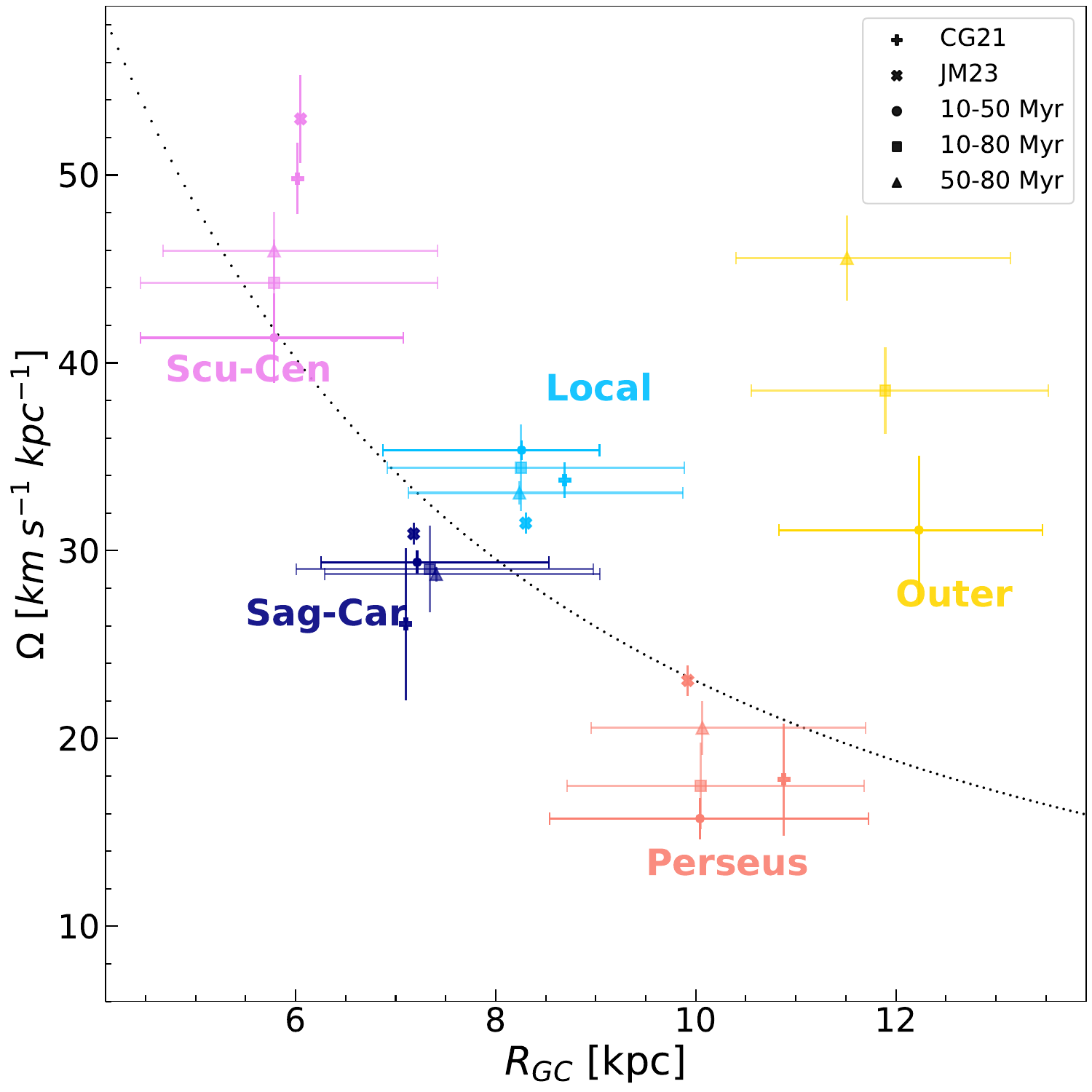}
    \caption{Pattern speed on different spiral arms, where the horizontal bars represent the range of OCs distributions on each spiral arm, the results from~\citet{Ginard21}~(10-50~Myr) and ~\citet{Joshi23}~(10-80~Myr) also included (without results for the Outer arm). 
In our analysis, we employed similar or identical parameters for R$_\odot$ and V$_\odot$ as in the above two studies.
The black dotted line in the figure represents the rotation curve from the Galactic gravitational potential model MWPotential2014 in $Galpy$~\citep{Bovy2013,Bovy2015}.The vertical bars represent the sample uncertainties calculated by the bootstrap method. The variation of Pattern speed caused by the uncertainty of other parameters are shown in Table 1.}
    \label{fig_pv}
\end{center}
\end{figure*}

\begin{table*}
\label{tab1}
\centering
\caption{\label{tab:example_table} The pattern speeds on different spiral arms, where "uncertainty type" refers to the error in fitting pattern speeds under uncertainties in sample, age, velocity, and distance. Additionally, the pattern speed with a bar potential~\citep{Dehnen99,Monari16} is presented, considering uncertainties only in the sample type.}
\vspace{1em} 
\begin{threeparttable}
    \begin{tabular}{lcccc} 
    \hline
    \hline
     Arm & $\Omega_{\text{p}}$ (10-50 Myr)& $\Omega_{\text{p}}$ (10-80 Myr) & $\Omega_{\text{p}}$ (50-80 Myr) &  Uncertainty Type \\
      & km~s$^{-1}$~kpc$^{-1}$ & km~s$^{-1}$~kpc$^{-1}$ & km~s$^{-1}$~kpc$^{-1}$ & \\
    \hline
    Scu-Cen & $43.17 \pm 2.25$& $47.13 \pm 2.43$ & $49.72 \pm 2.33$ &  Sample (with a bar potential)\\
    & $41.33 \pm 2.38$& $44.26 \pm 2.31$ & $45.98 \pm 2.08$ &  Sample\\
    & $41.19 \pm 0.81$ &$43.66 \pm 0.78$& $45.43 \pm 1.19$ & Age \\
    &  $40.85 \pm 0.47$ & $44.51 \pm 0.64$ & $46.19 \pm 0.10$ & Distance \\
    & $40.99 \pm 0.10$  & $44.88 \pm 0.38$ & $46.18 \pm 0.31$ & Radial velocity and proper motions \\
    \hline
    Sag-Car & $29.90 \pm 0.68$ & $29.63 \pm 0.33$ &  $29.46 \pm 0.35$ & Sample (with a bar potential) \\
    & $29.38 \pm 0.62$ & $29.01 \pm 0.28$ &  $28.75 \pm 0.38$ & Sample \\
     & $29.35 \pm 0.25$ & $28.99 \pm 0.15$ & $28.79 \pm 0.20$ & Age\\
     & $29.48 \pm 0.07$ & $29.08 \pm 0.04$ & $28.85 \pm 0.05$ &Distance \\
      & $29.46 \pm 0.06$ & $29.06 \pm 0.05$ & $28.87 \pm 0.06$ & Radial velocity and proper motions \\
     \hline
    Local & $35.85 \pm 0.53$ &$34.78 \pm 0.45$ & $33.57 \pm 0.50$ &  Sample (with a bar potential)\\
    &  $35.35 \pm 0.52$ &$34.41 \pm 0.45$ & $33.08 \pm 0.61$ &Sample \\
     & $35.41 \pm 0.21$ & $34.37 \pm 0.16$& $33.04 \pm 0.27$ & Age\\
     & $35.40 \pm 0.00$ & $34.38 \pm 0.04$ & $33.20 \pm 0.02$ &Distance \\
     &  $35.39 \pm 0.11$ & $34.45 \pm 0.10$ & $33.12 \pm 0.15$ & Radial velocity and proper motions  \\
     \hline
    Perseus & $15.89 \pm 1.10$& $17.59 \pm 1.30$ & $20.61 \pm 1.55$ &  Sample (with a bar potential)\\ 
     & $15.74 \pm 1.10$& $17.48 \pm 1.33$ & $20.56 \pm 1.42$ & Sample\\  
     &  $15.85 \pm 0.43$ & $17.47 \pm 0.45$ & $20.66 \pm 1.05$  &  Age \\
     &  $15.50 \pm 0.24$ & $17.43 \pm 0.28$ & $20.62 \pm 0.34$ &Distance \\
     & $15.41 \pm 0.12$ & $17.57 \pm 0.10 $ & $20.51 \pm 0.20$ &Radial velocity and proper motions  \\
     \hline
     Outer & $31.46 \pm 3.92$ & $38.59 \pm 2.32$&$45.94 \pm 2.37$ &  Sample (with a bar potential)\\
     & $31.10 \pm 3.97$ & $38.52 \pm 2.27$&$45.57 \pm 2.26$ &Sample\\
     & $31.78 \pm 2.33$  & $37.70 \pm 1.17$ & $45.17 \pm 2.44$& Age \\
     &  $31.24 \pm 1.57$ & $38.13 \pm 1.11$ & $46.31 \pm 1.08$ &Distance \\
     &  $30.96 \pm 0.37$ & $38.82 \pm 0.67$ & $46.31 \pm 0.15$ & Radial velocity and proper motions \\
    \hline
    \end{tabular}
\end{threeparttable}
\centering
\end{table*}

Our results (Figure~\ref{fig_pv}) show broad agreement with the findings of CG21 and JM23. 
We observed a general decrease in the pattern speeds of spiral arms with increasing distances from the Galactic center, with the exception of the Local arm and Outer arm. CG21 speculates that the anomalously high pattern speed of the Local arm may be attributed to its ongoing growing. Additionally, the limited number of clusters in the Outer arm, particularly near the Galactic disk's outer edge, suggests that the derived pattern speeds there may be overestimated due to insufficient OC sample. 
We find that the influence of the Galactic bar (Table 1) may slightly increase the pattern speed of the Scu-Cen arm by approximately 3 km~s$^{-1}$~kpc$^{-1}$. In contrast, the other four arms do not appear to be significantly affected by the bar. According to the quasi-stationary density wave theory~\citep{Shu16}, spiral arms maintain a constant pattern speed; however, the trends we present here and in previous studies (CG21, JM23) do not align with this prediction. Moreover, aside from the Outer arm, the pattern speeds of the spiral arms are consistent with the Galactic rotation curve, indicating a more dynamic structure for the arms. 

Accounting for errors, the Scu-Cen arm and Outer arm may share similar pattern speeds. If this is the case, the two arms located on the inner and outer sides of the Galactic disk could align with the predictions of the quasi-stationary density wave theory, which postulates that the arms have the same pattern speed. However, this possibility appears to be relatively low, as the CR is expected to be near the Scu-Cen arm in this context. Consequently, both the Sag-Car arm and the Perseus arm should display distinct age patterns, which are inconsistent with the observations presented in Section~\ref{sec4}. Additionally, the Outer arm's OCs are currently insufficient to delineate a clear age pattern for further validation of this scenario. Therefore, additional samples are required to accurately assess the pattern speed of the Outer arm.
In conclusion, it has been demonstrated that there is no significant age offset between the Local and Perseus arms, and the pattern speeds are consistent with the rotation curve. These findings are not compatible with the quasi-stationary density wave theory and support the consideration of a dynamic arm scenario for the Milky Way.

\section{Vertex deviation}
\label{sec5}

From another perspective, recent findings by~\citet{Funakoshi24} (hereafter F24) on classical Cepheids within the Perseus arm and Outer arm highlight patterns of disruption and growth respectively in these regions. Their approach revolved around analysing "vertex deviation ($lv$)", which is the inclination of velocity ellipsoid between Galactocentric radial motion and azimuthal motion ~\citep{Vorobyov08}. 
N-body simulations in F24, which align with the dynamic arm scenario featuring transient arms, indicated that a positive vertex deviation marks a disintegration phase in a spiral arm, whereas a negative deviation correlates with a growth phase. These findings are consistent with the relevance of vertex deviation as a sign of disruption, as discussed by \citet{Baba18}.

\begin{figure*}
\begin{center}
	\includegraphics[width=1.62\columnwidth]{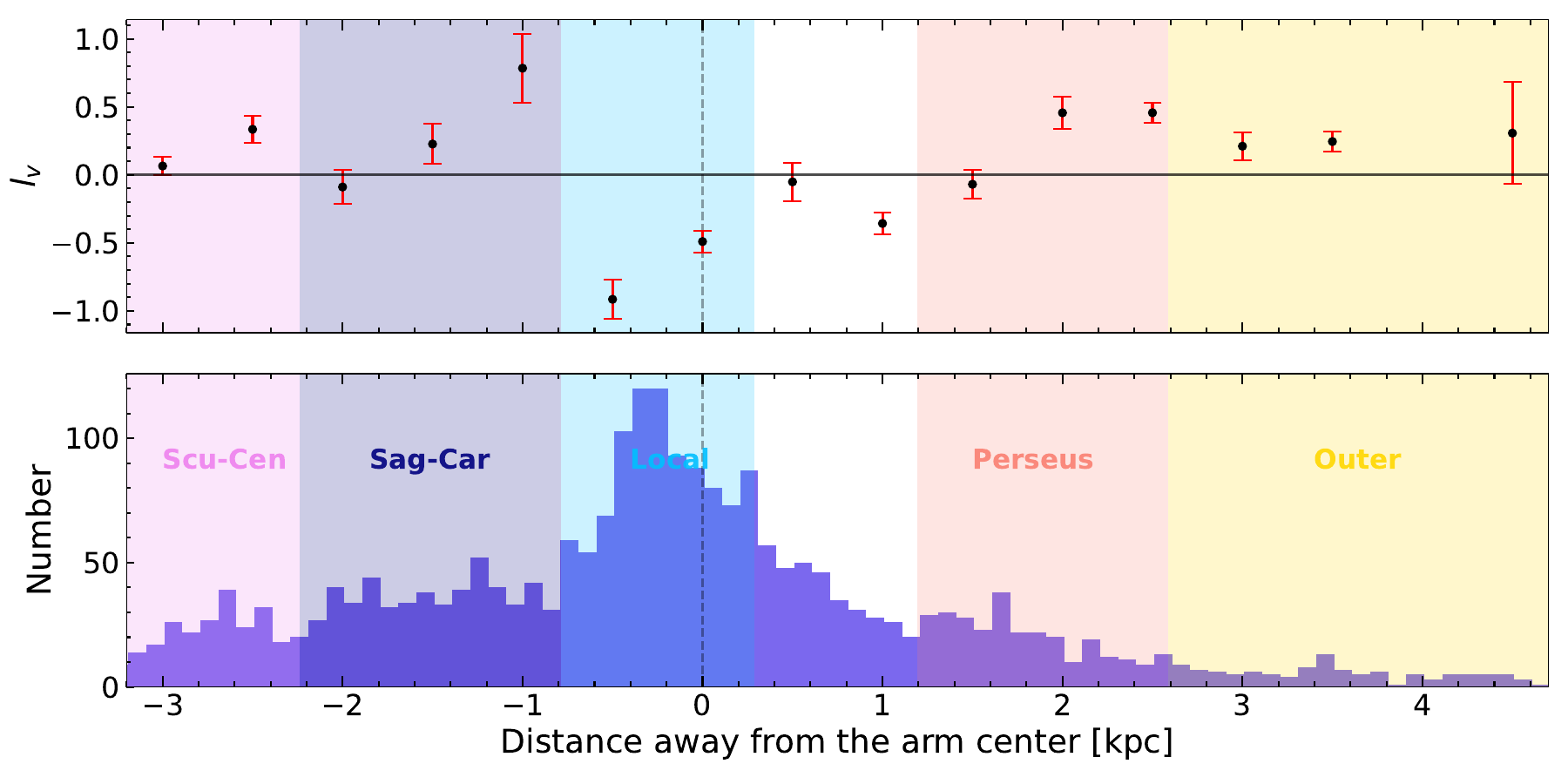}
    \caption{Upper panel: Distribution of vertex deviation for each spiral arm. The method used to calculate vertex deviations follows the approach outlined in F24 and has also been applied in other studies~\citep[e.g.][]{Zhao94,Simion21}. The selected bin interval range is 0.5~kpc, and intervals with cluster counts less than 20 are excluded. The $lv$ values and error bars are as follows: 70$\%$ of samples in each bin are randomly selected, this is repeated 1000 times, and then the mean and standard deviation of $lv$ are calculated from these 1000 iterations.
Lower panel: Histogram showing the distribution of OCs younger than 300~Myr; the x-axis represents the distance relative to the Local arm of~\citet{Reid19}.}
    \label{fig_lv}
\end{center}
\end{figure*}

Inspired by F24, we applied the same analytical model to calculate the vertex deviation for a sample of OCs younger than 300~Myr, aligned with the age range of the classical Cepheid samples. Utilizing the formula detailed in Section~2.2 of F24, our examination of $lv$ values across five spiral arms, as presented in Figure~\ref{fig_lv}, reveals predominantly smaller positive $lv$ values and zeroes in the Scu-Cen arm and Outer arm - with the latter having fewer members, which could impact the reliability of these results. Conversely, the Local arm registers significant negative values, with the nearby Sag-Car and Perseus arms exhibiting trends that are diametrically opposite. 
Interestingly, the inter-arm regions between the Local arm and the Perseus arm also show small negative $lv$ values, similar to those of the Local Arm.

In the dynamic spiral arms scenario, drawing on CG21, F24, and findings from our study, we conclude that the Local arm (and the inter-arm region between the Local and Perseus arms) is experiencing a phase of growth, a result consistent across the studies analyzing pattern speeds and vertex deviations. In contrast, both the Sag-Car arm and Perseus arm appear to be in states of disruption, aligning with F24's observations of the latter. This body of evidence supports prior research on spiral arm evolution.~\citet{Baba18} identified a similar pattern of disruption in the Perseus arm through Cepheid analysis, while recent work by~\citet{Asano24} traced motions across the Galactic disk, revealing growth trends in both the Local arm and Outer arm, and disruption in the Perseus arm. However, the Scu-Cen arm in the inner disk present weak signs of growth through pattern speed measurements and disruption signs from $lv$, it may be under the influence of the Galactic bar. Regarding the Outer arm, the results of F24 align with the conclusions of~\citet{Asano24}. However, in our study, the $lv$ values in OCs in the Outer arm do not exhibit the negative values noted in the Cepheids analyzed by F24. This discrepancy is likely attributable to the limited number of Outer disk clusters sampled in our research, underscoring the need for expanded studies with more extensive OC samples to clarify these results.


Within the context of dynamic arms, the arms are not static structures but rather transient features that arise due to gravitational instabilities or perturbations in the density of matter within the Galactic disk, and there is no substantial difference in rotational speeds between the material arms and the spiral pattern; they are effectively corotating across the Galaxy. In other words, the evolution of the arms does not give rise to a global age pattern throughout the entire Galaxy. However, these perturbations can also lead to the formation of density waves that propagate through the disk~\citep{Shu73,Goldreich78}, triggering variations in star formation activity along the arms~\citep{Roberts69}. Since the dynamical arm does not have a long-term persistence of the pattern, the resulting age distribution of stars within the arms can become randomized over time. Thus, the age gradient observed in the Sag-Car arm may be a random occurrence; it is also understandable why the observed offsets along the Sag-Car arm do not exhibit an increasing or decreasing trend with respect to the azimuthal angle $\phi$.

\section{Conclusion}
\label{sec6}

This paper represents a comprehensive investigation into the evolutionary trends of spiral arms in the Milky Way. Utilizing the Gaia DR3 catalogue of star clusters spanning various ages, we trace the structure of the spiral arms and find no consistent offsets among their positions. This challenges the predicted age pattern phenomenon associated with the quasi-stationary theory for spiral arms in the Milky Way. Through extensive dynamical simulations, we explore the evolution of the spiral arms by retracing the backward positions of star clusters.
We applied the same methodology as described in CG21 and F24 to calculate the pattern speeds and vertex deviations of spiral arms using OCs as tracers. Our key findings can be summarized as follows:

\begin{itemize}
\item[\textbullet] The pattern speeds of the spiral arms, apart from the Local  arm and Outer arm, decrease with increasing distance from the Galactic center. This trend corroborates the observations made in CG21 and JM23, underscoring the absence of a uniform pattern speed across the Milky Way's spiral arms.

\item[\textbullet]The presence of the bar increases the pattern speed of the Scu-Cen arm, while it has no significant effect on the other spiral arms further from the bar. The pattern speed of the Outer arm significantly exceeds the velocity of the rotation curve. This discrepancy may be attributed to the small sample size of OCs in the Outer arm, and further verification will require additional samples.

\item[\textbullet] In the case of the Local arm, a pronounced negative vertex deviation was observed, suggesting that this arm is currently expanding under the dynamic spiral arms scenario. Conversely, the Sag-Car arm and Perseus arm display trends that differ, with the observed patterns in the Perseus arm aligning with the results reported in F24.
\end{itemize}

Observational evidence increasingly challenges the applicability of the quasi-stationary theory to explain the origins of spiral arms in the Milky Way. This suggests that the spiral arms in the Milky Way, at least within the Galactic stellar disk (R$_{GC}$ < 15 kpc), do not conform to a grand-design pattern with a single pattern speed, but instead exhibit a multi-armed configuration. Since our OC sample is three times larger than that of the Gaia DR2 era, this conclusion reinforces the conclusion obtained by CG21 and JM23. However, the mechanisms driving the evolution of global patterns of the entire Galaxy remain elusive.

However, it is important to acknowledge that our study is limited by the sample size, preventing us from obtaining longer segments of spiral arms. At present, the details of the spiral arms observed in open star clusters only extend by 30 degrees. Especially for the Outer arm, which are based on a relatively small sample size. Therefore, further research is required to delve deeper into this complex issue and expand our understanding of spiral arm dynamics in the Milky Way. To fully understand the origin and evolution of age patterns within spiral arms, further observational studies, numerical simulations, and theoretical investigations are necessary. The complex interplay between various physical processes involved in spiral arm dynamics makes it challenging to discern the exact mechanisms responsible for the observed age distributions.

\section*{Acknowledgements}
We sincerely thank the referees for the insightful suggestions, which significantly enhanced the quality and impact of our paper.
This work was supported by National Natural Science Foundation of China through grants 12303024, the Natural Science Foundation of Sichuan Province (2024NSFSC0453), and the “Young Data Scientists” project of the National Astronomical Data Center (NADC2023YDS-07);
Y. Luo is supported by the NSFC under grant 12173028, the CSST project: CMS-CSST-2021-A10; K. Wang is supported by the NSFC 12373035.
And this work has made use of data from the European Space Agency (ESA) mission GAIA (\url{https://www.cosmos.esa.int/gaia}), processed by the GAIA Data Processing and Analysis Consortium (DPAC,\url{https://www.cosmos.esa.int/web/gaia/dpac/consortium}). Funding for the DPAC has been provided by national institutions, in particular the institutions participating in the GAIA Multilateral Agreement.

\section*{Data Availability}
The open cluster data used in this article and the simulation snapshots will be shared through a request to the corresponding author.



\bibliographystyle{mnras}
\bibliography{arm} 

\begin{thebibliography}{}
\makeatletter
\relax
\def\mn@urlcharsother{\let\do\@makeother \do\$\do\&\do\#\do\^\do\_\do\%\do\~}
\def\mn@doi{\begingroup\mn@urlcharsother \@ifnextchar [ {\mn@doi@}
  {\mn@doi@[]}}
\def\mn@doi@[#1]#2{\def\@tempa{#1}\ifx\@tempa\@empty \href
  {http://dx.doi.org/#2} {doi:#2}\else \href {http://dx.doi.org/#2} {#1}\fi
  \endgroup}
\def\mn@eprint#1#2{\mn@eprint@#1:#2::\@nil}
\def\mn@eprint@arXiv#1{\href {http://arxiv.org/abs/#1} {{\tt arXiv:#1}}}
\def\mn@eprint@dblp#1{\href {http://dblp.uni-trier.de/rec/bibtex/#1.xml}
  {dblp:#1}}
\def\mn@eprint@#1:#2:#3:#4\@nil{\def\@tempa {#1}\def\@tempb {#2}\def\@tempc
  {#3}\ifx \@tempc \@empty \let \@tempc \@tempb \let \@tempb \@tempa \fi \ifx
  \@tempb \@empty \def\@tempb {arXiv}\fi \@ifundefined
  {mn@eprint@\@tempb}{\@tempb:\@tempc}{\expandafter \expandafter \csname
  mn@eprint@\@tempb\endcsname \expandafter{\@tempc}}}

\bibitem[\protect\citeauthoryear{{Am{\^o}res}, {L{\'e}pine}  \&
  {Mishurov}}{{Am{\^o}res} et~al.}{2009}]{Amores09}
{Am{\^o}res} E.~B.,  {L{\'e}pine} J.~R.~D.,   {Mishurov} Y.~N.,  2009, \mn@doi
  [\mnras] {10.1111/j.1365-2966.2009.15611.x}, \href
  {https://ui.adsabs.harvard.edu/abs/2009MNRAS.400.1768A} {400, 1768}

\bibitem[\protect\citeauthoryear{{Asano}, {Kawata}, {Fujii}  \& {Baba}}{{Asano}
  et~al.}{2024}]{Asano24}
{Asano} T.,  {Kawata} D.,  {Fujii} M.~S.,   {Baba} J.,  2024, \mn@doi [\mnras]
  {10.1093/mnrasl/slad190}, \href
  {https://ui.adsabs.harvard.edu/abs/2024MNRAS.529L...7A} {529, L7}

\bibitem[\protect\citeauthoryear{{Baba}, {Kawata}, {Matsunaga}, {Grand}  \&
  {Hunt}}{{Baba} et~al.}{2018}]{Baba18}
{Baba} J.,  {Kawata} D.,  {Matsunaga} N.,  {Grand} R. J.~J.,   {Hunt} J. A.~S.,
   2018, \mn@doi [\apjl] {10.3847/2041-8213/aaa839}, \href
  {https://ui.adsabs.harvard.edu/abs/2018ApJ...853L..23B} {853, L23}

\bibitem[\protect\citeauthoryear{{Bovy}}{{Bovy}}{2015}]{Bovy2015}
{Bovy} J.,  2015, \mn@doi [\apjs] {10.1088/0067-0049/216/2/29}, \href
  {https://ui.adsabs.harvard.edu/abs/2015ApJS..216...29B} {216, 29}

\bibitem[\protect\citeauthoryear{{Bovy} \& {Rix}}{{Bovy} \&
  {Rix}}{2013}]{Bovy2013}
{Bovy} J.,  {Rix} H.-W.,  2013, \mn@doi [\apj] {10.1088/0004-637X/779/2/115},
  \href {https://ui.adsabs.harvard.edu/abs/2013ApJ...779..115B} {779, 115}

\bibitem[\protect\citeauthoryear{{Bressan}, {Marigo}, {Girardi}, {Salasnich},
  {Dal Cero}, {Rubele}  \& {Nanni}}{{Bressan} et~al.}{2012}]{Bressan12}
{Bressan} A.,  {Marigo} P.,  {Girardi} L.,  {Salasnich} B.,  {Dal Cero} C.,
  {Rubele} S.,   {Nanni} A.,  2012, \mn@doi [\mnras]
  {10.1111/j.1365-2966.2012.21948.x}, \href
  {https://ui.adsabs.harvard.edu/abs/2012MNRAS.427..127B} {427, 127}

\bibitem[\protect\citeauthoryear{{Cantat-Gaudin} et~al.,}{{Cantat-Gaudin}
  et~al.}{2018}]{Gaudin18}
{Cantat-Gaudin} T.,  et~al., 2018, \mn@doi [\aap]
  {10.1051/0004-6361/201833476}, \href
  {https://ui.adsabs.harvard.edu/abs/2018A&A...618A..93C} {618, A93}

\bibitem[\protect\citeauthoryear{{Cantat-Gaudin} et~al.,}{{Cantat-Gaudin}
  et~al.}{2020}]{Gaudin20a}
{Cantat-Gaudin} T.,  et~al., 2020, \mn@doi [\aap]
  {10.1051/0004-6361/202038192}, \href
  {https://ui.adsabs.harvard.edu/abs/2020A&A...640A...1C} {640, A1}

\bibitem[\protect\citeauthoryear{{Castro-Ginard} et~al.,}{{Castro-Ginard}
  et~al.}{2020}]{Ginard20}
{Castro-Ginard} A.,  et~al., 2020, \mn@doi [\aap]
  {10.1051/0004-6361/201937386}, \href
  {https://ui.adsabs.harvard.edu/abs/2020A&A...635A..45C} {635, A45}

\bibitem[\protect\citeauthoryear{{Castro-Ginard} et~al.,}{{Castro-Ginard}
  et~al.}{2021}]{Ginard21}
{Castro-Ginard} A.,  et~al., 2021, \mn@doi [\aap]
  {10.1051/0004-6361/202039751}, \href
  {https://ui.adsabs.harvard.edu/abs/2021A&A...652A.162C} {652, A162}

\bibitem[\protect\citeauthoryear{{Castro-Ginard} et~al.,}{{Castro-Ginard}
  et~al.}{2022}]{Ginard22}
{Castro-Ginard} A.,  et~al., 2022, \mn@doi [\aap]
  {10.1051/0004-6361/202142568}, \href
  {https://ui.adsabs.harvard.edu/abs/2022A&A...661A.118C} {661, A118}

\bibitem[\protect\citeauthoryear{{Cheng}, {Liu}, {Mao}  \& {Cui}}{{Cheng}
  et~al.}{2019}]{Cheng19}
{Cheng} X.,  {Liu} C.,  {Mao} S.,   {Cui} W.,  2019, \mn@doi [\apjl]
  {10.3847/2041-8213/ab020e}, \href
  {https://ui.adsabs.harvard.edu/abs/2019ApJ...872L...1C} {872, L1}

\bibitem[\protect\citeauthoryear{{Choi}, {Dalcanton}, {Williams}, {Weisz},
  {Skillman}, {Fouesneau}  \& {Dolphin}}{{Choi} et~al.}{2015}]{Choi15}
{Choi} Y.,  {Dalcanton} J.~J.,  {Williams} B.~F.,  {Weisz} D.~R.,  {Skillman}
  E.~D.,  {Fouesneau} M.,   {Dolphin} A.~E.,  2015, \mn@doi [\apj]
  {10.1088/0004-637X/810/1/9}, \href
  {https://ui.adsabs.harvard.edu/abs/2015ApJ...810....9C} {810, 9}

\bibitem[\protect\citeauthoryear{{Cohen}, {Cong}, {Dame}  \&
  {Thaddeus}}{{Cohen} et~al.}{1980}]{Cohen80}
{Cohen} R.~S.,  {Cong} H.,  {Dame} T.~M.,   {Thaddeus} P.,  1980, \mn@doi
  [\apjl] {10.1086/183290}, \href
  {https://ui.adsabs.harvard.edu/abs/1980ApJ...239L..53C} {239, L53}

\bibitem[\protect\citeauthoryear{{Dame} \& {Thaddeus}}{{Dame} \&
  {Thaddeus}}{2011}]{Dame11}
{Dame} T.~M.,  {Thaddeus} P.,  2011, \mn@doi [\apjl]
  {10.1088/2041-8205/734/1/L24}, \href
  {https://ui.adsabs.harvard.edu/abs/2011ApJ...734L..24D} {734, L24}

\bibitem[\protect\citeauthoryear{{Dehnen}}{{Dehnen}}{1999}]{Dehnen99}
{Dehnen} W.,  1999, \mn@doi [\apjl] {10.1086/312299}, \href
  {https://ui.adsabs.harvard.edu/abs/1999ApJ...524L..35D} {524, L35}

\bibitem[\protect\citeauthoryear{{Dias} \& {L{\'e}pine}}{{Dias} \&
  {L{\'e}pine}}{2005}]{Dias05}
{Dias} W.~S.,  {L{\'e}pine} J.~R.~D.,  2005, \mn@doi [\apj] {10.1086/431456},
  \href {https://ui.adsabs.harvard.edu/abs/2005ApJ...629..825D} {629, 825}

\bibitem[\protect\citeauthoryear{{Dias}, {Monteiro}, {L{\'e}pine}  \&
  {Barros}}{{Dias} et~al.}{2019}]{Dias19}
{Dias} W.~S.,  {Monteiro} H.,  {L{\'e}pine} J.~R.~D.,   {Barros} D.~A.,  2019,
  \mn@doi [\mnras] {10.1093/mnras/stz1196}, \href
  {https://ui.adsabs.harvard.edu/abs/2019MNRAS.486.5726D} {486, 5726}

\bibitem[\protect\citeauthoryear{{Dobbs} \& {Baba}}{{Dobbs} \&
  {Baba}}{2014}]{Dobbs14}
{Dobbs} C.,  {Baba} J.,  2014, \mn@doi [\pasa] {10.1017/pasa.2014.31}, \href
  {https://ui.adsabs.harvard.edu/abs/2014PASA...31...35D} {31, e035}

\bibitem[\protect\citeauthoryear{{Dobbs} \& {Pringle}}{{Dobbs} \&
  {Pringle}}{2010}]{Dobbs10b}
{Dobbs} C.~L.,  {Pringle} J.~E.,  2010, \mn@doi [\mnras]
  {10.1111/j.1365-2966.2010.17323.x}, \href
  {https://ui.adsabs.harvard.edu/abs/2010MNRAS.409..396D} {409, 396}

\bibitem[\protect\citeauthoryear{{Funakoshi}, {Matsunaga}, {Kawata}, {Baba},
  {Taniguchi}  \& {Fujii}}{{Funakoshi} et~al.}{2024}]{Funakoshi24}
{Funakoshi} N.,  {Matsunaga} N.,  {Kawata} D.,  {Baba} J.,  {Taniguchi} D.,
  {Fujii} M.,  2024, \mn@doi [arXiv e-prints] {10.48550/arXiv.2401.13037},
  \href {https://ui.adsabs.harvard.edu/abs/2024arXiv240113037F} {p.
  arXiv:2401.13037}

\bibitem[\protect\citeauthoryear{{Gaia Collaboration} et~al.,}{{Gaia
  Collaboration} et~al.}{2018}]{Collaboration18}
{Gaia Collaboration} et~al., 2018, \mn@doi [\aap]
  {10.1051/0004-6361/201833051}, \href
  {https://ui.adsabs.harvard.edu/abs/2018A&A...616A...1G} {616, A1}

\bibitem[\protect\citeauthoryear{{Gaia Collaboration} et~al.,}{{Gaia
  Collaboration} et~al.}{2021}]{Collaboration21}
{Gaia Collaboration} et~al., 2021, \mn@doi [\aap]
  {10.1051/0004-6361/202039657}, \href
  {https://ui.adsabs.harvard.edu/abs/2021A&A...649A...1G} {649, A1}

\bibitem[\protect\citeauthoryear{{Gaia Collaboration} et~al.,}{{Gaia
  Collaboration} et~al.}{2023}]{Collaboration23}
{Gaia Collaboration} et~al., 2023, \mn@doi [\aap]
  {10.1051/0004-6361/202243797}, \href
  {https://ui.adsabs.harvard.edu/abs/2023A&A...674A..37G} {674, A37}

\bibitem[\protect\citeauthoryear{{Garner}, {Mihos}, {Harding}  \&
  {Garner}}{{Garner} et~al.}{2024}]{Garner24}
{Garner} R.,  {Mihos} J.~C.,  {Harding} P.,   {Garner} C.~R.,  2024, \mn@doi
  [\apj] {10.3847/1538-4357/ad0e63}, \href
  {https://ui.adsabs.harvard.edu/abs/2024ApJ...961..217G} {961, 217}

\bibitem[\protect\citeauthoryear{{Georgelin} \& {Georgelin}}{{Georgelin} \&
  {Georgelin}}{1976}]{Georgelin76}
{Georgelin} Y.~M.,  {Georgelin} Y.~P.,  1976, \aap, \href
  {https://ui.adsabs.harvard.edu/abs/1976A&A....49...57G} {49, 57}

\bibitem[\protect\citeauthoryear{{Goldreich} \& {Tremaine}}{{Goldreich} \&
  {Tremaine}}{1978}]{Goldreich78}
{Goldreich} P.,  {Tremaine} S.,  1978, \mn@doi [\apj] {10.1086/156203}, \href
  {https://ui.adsabs.harvard.edu/abs/1978ApJ...222..850G} {222, 850}

\bibitem[\protect\citeauthoryear{{He}}{{He}}{2023}]{He23a}
{He} Z.,  2023, \mn@doi [\apjl] {10.3847/2041-8213/ace77d}, \href
  {https://ui.adsabs.harvard.edu/abs/2023ApJ...954L...9H} {954, L9}

\bibitem[\protect\citeauthoryear{{He}, {Xu}  \& {Hou}}{{He}
  et~al.}{2021a}]{He21b}
{He} Z.-H.,  {Xu} Y.,   {Hou} L.-G.,  2021a, \mn@doi [Research in Astronomy and
  Astrophysics] {10.1088/1674-4527/21/1/9}, \href
  {https://ui.adsabs.harvard.edu/abs/2021RAA....21....9H} {21, 009}

\bibitem[\protect\citeauthoryear{{He}, {Xu}, {Hao}, {Wu}  \& {Li}}{{He}
  et~al.}{2021b}]{He21a}
{He} Z.-H.,  {Xu} Y.,  {Hao} C.-J.,  {Wu} Z.-Y.,   {Li} J.-J.,  2021b, \mn@doi
  [Research in Astronomy and Astrophysics] {10.1088/1674-4527/21/4/93}, \href
  {https://ui.adsabs.harvard.edu/abs/2021RAA....21...93H} {21, 093}

\bibitem[\protect\citeauthoryear{{He} et~al.,}{{He} et~al.}{2022a}]{He2022a}
{He} Z.,  et~al., 2022a, \mn@doi [\apjs] {10.3847/1538-4365/ac5cbb}, \href
  {https://ui.adsabs.harvard.edu/abs/2022ApJS..260....8H} {260, 8}

\bibitem[\protect\citeauthoryear{{He}, {Wang}, {Luo}, {Li}, {Liu}  \&
  {Jiang}}{{He} et~al.}{2022b}]{He22b}
{He} Z.,  {Wang} K.,  {Luo} Y.,  {Li} J.,  {Liu} X.,   {Jiang} Q.,  2022b,
  \mn@doi [\apjs] {10.3847/1538-4365/ac7c17}, \href
  {https://ui.adsabs.harvard.edu/abs/2022ApJS..262....7H} {262, 7}

\bibitem[\protect\citeauthoryear{{He}, {Liu}, {Luo}, {Wang}  \& {Jiang}}{{He}
  et~al.}{2023a}]{He2023a}
{He} Z.,  {Liu} X.,  {Luo} Y.,  {Wang} K.,   {Jiang} Q.,  2023a, \mn@doi
  [\apjs] {10.3847/1538-4365/ac9af8}, \href
  {https://ui.adsabs.harvard.edu/abs/2023ApJS..264....8H} {264, 8}

\bibitem[\protect\citeauthoryear{{He}, {Luo}, {Wang}, {Ren}, {Peng}, {Cui},
  {Liu}  \& {Jiang}}{{He} et~al.}{2023b}]{He23b}
{He} Z.,  {Luo} Y.,  {Wang} K.,  {Ren} A.,  {Peng} L.,  {Cui} Q.,  {Liu} X.,
  {Jiang} Q.,  2023b, \mn@doi [\apjs] {10.3847/1538-4365/acd6fa}, \href
  {https://ui.adsabs.harvard.edu/abs/2023ApJS..267...34H} {267, 34}

\bibitem[\protect\citeauthoryear{{Hou} \& {Han}}{{Hou} \& {Han}}{2014}]{Hou14}
{Hou} L.~G.,  {Han} J.~L.,  2014, \mn@doi [\aap] {10.1051/0004-6361/201424039},
  \href {https://ui.adsabs.harvard.edu/abs/2014A&A...569A.125H} {569, A125}

\bibitem[\protect\citeauthoryear{{Hou} \& {Han}}{{Hou} \& {Han}}{2015}]{Hou15}
{Hou} L.~G.,  {Han} J.~L.,  2015, \mn@doi [\mnras] {10.1093/mnras/stv1904},
  \href {https://ui.adsabs.harvard.edu/abs/2015MNRAS.454..626H} {454, 626}

\bibitem[\protect\citeauthoryear{{Hunt} \& {Reffert}}{{Hunt} \&
  {Reffert}}{2021}]{Hunt21}
{Hunt} E.~L.,  {Reffert} S.,  2021, \mn@doi [\aap]
  {10.1051/0004-6361/202039341}, \href
  {https://ui.adsabs.harvard.edu/abs/2021A&A...646A.104H} {646, A104}

\bibitem[\protect\citeauthoryear{{Hunt} \& {Reffert}}{{Hunt} \&
  {Reffert}}{2023}]{Hunt23}
{Hunt} E.~L.,  {Reffert} S.,  2023, \mn@doi [\aap]
  {10.1051/0004-6361/202346285}, \href
  {https://ui.adsabs.harvard.edu/abs/2023A&A...673A.114H} {673, A114}

\bibitem[\protect\citeauthoryear{{Joshi} \& {Malhotra}}{{Joshi} \&
  {Malhotra}}{2023}]{Joshi23}
{Joshi} Y.~C.,  {Malhotra} S.,  2023, \mn@doi [\aj] {10.3847/1538-3881/acf7c8},
  \href {https://ui.adsabs.harvard.edu/abs/2023AJ....166..170J} {166, 170}

\bibitem[\protect\citeauthoryear{{Katz} et~al.,}{{Katz} et~al.}{2023}]{gaia23}
{Katz} D.,  et~al., 2023, \mn@doi [\aap] {10.1051/0004-6361/202244220}, \href
  {https://ui.adsabs.harvard.edu/abs/2023A&A...674A...5K} {674, A5}

\bibitem[\protect\citeauthoryear{{Kounkel}, {Covey}  \& {Stassun}}{{Kounkel}
  et~al.}{2020}]{Kounkel20}
{Kounkel} M.,  {Covey} K.,   {Stassun} K.~G.,  2020, \mn@doi [\aj]
  {10.3847/1538-3881/abc0e6}, \href
  {https://ui.adsabs.harvard.edu/abs/2020AJ....160..279K} {160, 279}

\bibitem[\protect\citeauthoryear{{Levine}, {Blitz}  \& {Heiles}}{{Levine}
  et~al.}{2006}]{Levine06}
{Levine} E.~S.,  {Blitz} L.,   {Heiles} C.,  2006, \mn@doi [Science]
  {10.1126/science.1128455}, \href
  {https://ui.adsabs.harvard.edu/abs/2006Sci...312.1773L} {312, 1773}

\bibitem[\protect\citeauthoryear{{Lin} \& {Shu}}{{Lin} \& {Shu}}{1964}]{Lin64}
{Lin} C.~C.,  {Shu} F.~H.,  1964, \mn@doi [\apj] {10.1086/147955}, \href
  {https://ui.adsabs.harvard.edu/abs/1964ApJ...140..646L} {140, 646}

\bibitem[\protect\citeauthoryear{{Lin} \& {Shu}}{{Lin} \& {Shu}}{1966}]{Lin66}
{Lin} C.~C.,  {Shu} F.~H.,  1966, \mn@doi [Proceedings of the National Academy
  of Science] {10.1073/pnas.55.2.229}, \href
  {https://ui.adsabs.harvard.edu/abs/1966PNAS...55..229L} {55, 229}

\bibitem[\protect\citeauthoryear{{Lin}, {Xu}, {Hou}, {Liu}, {Li}, {Hao}, {Li}
  \& {Bian}}{{Lin} et~al.}{2022}]{Lin22}
{Lin} Z.,  {Xu} Y.,  {Hou} L.,  {Liu} D.,  {Li} Y.,  {Hao} C.,  {Li} J.,
  {Bian} S.,  2022, \mn@doi [\apj] {10.3847/1538-4357/ac67a6}, \href
  {https://ui.adsabs.harvard.edu/abs/2022ApJ...931...72L} {931, 72}

\bibitem[\protect\citeauthoryear{{Liu} \& {Pang}}{{Liu} \&
  {Pang}}{2019}]{Liu19}
{Liu} L.,  {Pang} X.,  2019, \mn@doi [\apjs] {10.3847/1538-4365/ab530a}, \href
  {https://ui.adsabs.harvard.edu/abs/2019ApJS..245...32L} {245, 32}

\bibitem[\protect\citeauthoryear{{Michtchenko}, {L{\'e}pine},
  {P{\'e}rez-Villegas}, {Vieira}  \& {Barros}}{{Michtchenko}
  et~al.}{2018}]{Michtchenko18}
{Michtchenko} T.~A.,  {L{\'e}pine} J. R.~D.,  {P{\'e}rez-Villegas} A.,
  {Vieira} R. S.~S.,   {Barros} D.~A.,  2018, \mn@doi [\apjl]
  {10.3847/2041-8213/aad804}, \href
  {https://ui.adsabs.harvard.edu/abs/2018ApJ...863L..37M} {863, L37}

\bibitem[\protect\citeauthoryear{{Monari}, {Famaey}, {Siebert}, {Grand},
  {Kawata}  \& {Boily}}{{Monari} et~al.}{2016}]{Monari16}
{Monari} G.,  {Famaey} B.,  {Siebert} A.,  {Grand} R. J.~J.,  {Kawata} D.,
  {Boily} C.,  2016, \mn@doi [\mnras] {10.1093/mnras/stw1564}, \href
  {https://ui.adsabs.harvard.edu/abs/2016MNRAS.461.3835M} {461, 3835}

\bibitem[\protect\citeauthoryear{{Morgan}, {Whitford}  \& {Code}}{{Morgan}
  et~al.}{1953}]{Morgan53}
{Morgan} W.~W.,  {Whitford} A.~E.,   {Code} A.~D.,  1953, \mn@doi [\apj]
  {10.1086/145754}, \href
  {https://ui.adsabs.harvard.edu/abs/1953ApJ...118..318M} {118, 318}

\bibitem[\protect\citeauthoryear{{Peterken}, {Merrifield},
  {Arag{\'o}n-Salamanca}, {Drory}, {Krawczyk}, {Masters}, {Weijmans}  \&
  {Westfall}}{{Peterken} et~al.}{2019}]{Peterken19}
{Peterken} T.~G.,  {Merrifield} M.~R.,  {Arag{\'o}n-Salamanca} A.,  {Drory} N.,
   {Krawczyk} C.~M.,  {Masters} K.~L.,  {Weijmans} A.-M.,   {Westfall} K.~B.,
  2019, \mn@doi [Nature Astronomy] {10.1038/s41550-018-0627-5}, \href
  {https://ui.adsabs.harvard.edu/abs/2019NatAs...3..178P} {3, 178}

\bibitem[\protect\citeauthoryear{{Reid} et~al.,}{{Reid} et~al.}{2019}]{Reid19}
{Reid} M.~J.,  et~al., 2019, \mn@doi [\apj] {10.3847/1538-4357/ab4a11}, \href
  {https://ui.adsabs.harvard.edu/abs/2019ApJ...885..131R} {885, 131}

\bibitem[\protect\citeauthoryear{{Roberts}}{{Roberts}}{1969}]{Roberts69}
{Roberts} W.~W.,  1969, \mn@doi [\apj] {10.1086/150177}, \href
  {https://ui.adsabs.harvard.edu/abs/1969ApJ...158..123R} {158, 123}

\bibitem[\protect\citeauthoryear{{Sellwood}}{{Sellwood}}{2011}]{Sellwood11}
{Sellwood} J.~A.,  2011, \mn@doi [\mnras] {10.1111/j.1365-2966.2010.17545.x},
  \href {https://ui.adsabs.harvard.edu/abs/2011MNRAS.410.1637S} {410, 1637}

\bibitem[\protect\citeauthoryear{{Sellwood} \& {Carlberg}}{{Sellwood} \&
  {Carlberg}}{1984}]{Sellwood84}
{Sellwood} J.~A.,  {Carlberg} R.~G.,  1984, \mn@doi [\apj] {10.1086/162176},
  \href {https://ui.adsabs.harvard.edu/abs/1984ApJ...282...61S} {282, 61}

\bibitem[\protect\citeauthoryear{{Sellwood} \& {Carlberg}}{{Sellwood} \&
  {Carlberg}}{2014}]{Sellwood14}
{Sellwood} J.~A.,  {Carlberg} R.~G.,  2014, \mn@doi [\apj]
  {10.1088/0004-637X/785/2/137}, \href
  {https://ui.adsabs.harvard.edu/abs/2014ApJ...785..137S} {785, 137}

\bibitem[\protect\citeauthoryear{{Sellwood} \& {Carlberg}}{{Sellwood} \&
  {Carlberg}}{2019}]{Sellwood19b}
{Sellwood} J.~A.,  {Carlberg} R.~G.,  2019, \mn@doi [\mnras]
  {10.1093/mnras/stz2132}, \href
  {https://ui.adsabs.harvard.edu/abs/2019MNRAS.489..116S} {489, 116}

\bibitem[\protect\citeauthoryear{{Shabani} et~al.,}{{Shabani}
  et~al.}{2018}]{Shabani18}
{Shabani} F.,  et~al., 2018, \mn@doi [\mnras] {10.1093/mnras/sty1277}, \href
  {https://ui.adsabs.harvard.edu/abs/2018MNRAS.478.3590S} {478, 3590}

\bibitem[\protect\citeauthoryear{{Shu}}{{Shu}}{2016}]{Shu16}
{Shu} F.~H.,  2016, \mn@doi [\araa] {10.1146/annurev-astro-081915-023426},
  \href {https://ui.adsabs.harvard.edu/abs/2016ARA&A..54..667S} {54, 667}

\bibitem[\protect\citeauthoryear{{Shu}, {Milione}  \& {Roberts}}{{Shu}
  et~al.}{1973}]{Shu73}
{Shu} F.~H.,  {Milione} V.,   {Roberts} William~W. J.,  1973, \mn@doi [\apj]
  {10.1086/152270}, \href
  {https://ui.adsabs.harvard.edu/abs/1973ApJ...183..819S} {183, 819}

\bibitem[\protect\citeauthoryear{{Sim}, {Lee}, {Ann}  \& {Kim}}{{Sim}
  et~al.}{2019}]{Sim19}
{Sim} G.,  {Lee} S.~H.,  {Ann} H.~B.,   {Kim} S.,  2019, \mn@doi [Journal of
  Korean Astronomical Society] {10.5303/JKAS.2019.52.5.145}, \href
  {https://ui.adsabs.harvard.edu/abs/2019JKAS...52..145S} {52, 145}

\bibitem[\protect\citeauthoryear{{Simion}, {Shen}, {Koposov}, {Ness},
  {Freeman}, {Bland-Hawthorn}  \& {Lewis}}{{Simion} et~al.}{2021}]{Simion21}
{Simion} I.~T.,  {Shen} J.,  {Koposov} S.~E.,  {Ness} M.,  {Freeman} K.,
  {Bland-Hawthorn} J.,   {Lewis} G.~F.,  2021, \mn@doi [\mnras]
  {10.1093/mnras/stab073}, \href
  {https://ui.adsabs.harvard.edu/abs/2021MNRAS.502.1740S} {502, 1740}

\bibitem[\protect\citeauthoryear{{Skowron} et~al.,}{{Skowron}
  et~al.}{2019}]{Skowron19}
{Skowron} D.~M.,  et~al., 2019, \mn@doi [Science] {10.1126/science.aau3181},
  \href {https://ui.adsabs.harvard.edu/abs/2019Sci...365..478S} {365, 478}

\bibitem[\protect\citeauthoryear{{Sun}, {Xu}, {Yang}, {Li}, {Du}, {Zhang}  \&
  {Zhou}}{{Sun} et~al.}{2015}]{Sun15}
{Sun} Y.,  {Xu} Y.,  {Yang} J.,  {Li} F.-C.,  {Du} X.-Y.,  {Zhang} S.-B.,
  {Zhou} X.,  2015, \mn@doi [\apjl] {10.1088/2041-8205/798/2/L27}, \href
  {https://ui.adsabs.harvard.edu/abs/2015ApJ...798L..27S} {798, L27}

\bibitem[\protect\citeauthoryear{{Veselova} \& {Nikiforov}}{{Veselova} \&
  {Nikiforov}}{2020}]{Veselova20}
{Veselova} A.~V.,  {Nikiforov} I.,  2020, \mn@doi [Research in Astronomy and
  Astrophysics] {10.1088/1674-4527/20/12/209}, \href
  {https://ui.adsabs.harvard.edu/abs/2020RAA....20..209V} {20, 209}

\bibitem[\protect\citeauthoryear{{Vorobyov} \& {Theis}}{{Vorobyov} \&
  {Theis}}{2008}]{Vorobyov08}
{Vorobyov} E.~I.,  {Theis} C.,  2008, \mn@doi [\mnras]
  {10.1111/j.1365-2966.2007.12476.x}, \href
  {https://ui.adsabs.harvard.edu/abs/2008MNRAS.383..817V} {383, 817}

\bibitem[\protect\citeauthoryear{{Wada}, {Baba}  \& {Saitoh}}{{Wada}
  et~al.}{2011}]{Wada11}
{Wada} K.,  {Baba} J.,   {Saitoh} T.~R.,  2011, \mn@doi [\apj]
  {10.1088/0004-637X/735/1/1}, \href
  {https://ui.adsabs.harvard.edu/abs/2011ApJ...735....1W} {735, 1}

\bibitem[\protect\citeauthoryear{{Xu} et~al.,}{{Xu} et~al.}{2016}]{Xu16}
{Xu} Y.,  et~al., 2016, \mn@doi [Science Advances] {10.1126/sciadv.1600878},
  \href {https://ui.adsabs.harvard.edu/abs/2016SciA....2E0878X} {2, e1600878}

\bibitem[\protect\citeauthoryear{{Xu} et~al.,}{{Xu} et~al.}{2018}]{Xu18}
{Xu} Y.,  et~al., 2018, \mn@doi [\aap] {10.1051/0004-6361/201833407}, \href
  {https://ui.adsabs.harvard.edu/abs/2018A&A...616L..15X} {616, L15}

\bibitem[\protect\citeauthoryear{{Xu}, {Hao}, {Liu}, {Lin}, {Bian}, {Hou}, {Li}
   \& {Li}}{{Xu} et~al.}{2023}]{Xu23}
{Xu} Y.,  {Hao} C.~J.,  {Liu} D.~J.,  {Lin} Z.~H.,  {Bian} S.~B.,  {Hou} L.~G.,
   {Li} J.~J.,   {Li} Y.~J.,  2023, \mn@doi [\apj] {10.3847/1538-4357/acc45c},
  \href {https://ui.adsabs.harvard.edu/abs/2023ApJ...947...54X} {947, 54}

\bibitem[\protect\citeauthoryear{{Yu} \& {Ho}}{{Yu} \& {Ho}}{2018}]{YuYu18}
{Yu} S.-Y.,  {Ho} L.~C.,  2018, \mn@doi [\apj] {10.3847/1538-4357/aaeacd},
  \href {https://ui.adsabs.harvard.edu/abs/2018ApJ...869...29Y} {869, 29}

\bibitem[\protect\citeauthoryear{{Zhao}, {Spergel}  \& {Rich}}{{Zhao}
  et~al.}{1994}]{Zhao94}
{Zhao} H.,  {Spergel} D.~N.,   {Rich} R.~M.,  1994, \mn@doi [\aj]
  {10.1086/117227}, \href
  {https://ui.adsabs.harvard.edu/abs/1994AJ....108.2154Z} {108, 2154}

\makeatother
\end{thebibliography}




\appendix
\section{Sketch map of age pattern}
\label{appendix}
\begin{figure*}
\begin{center}
	\includegraphics[width=1.32\columnwidth]{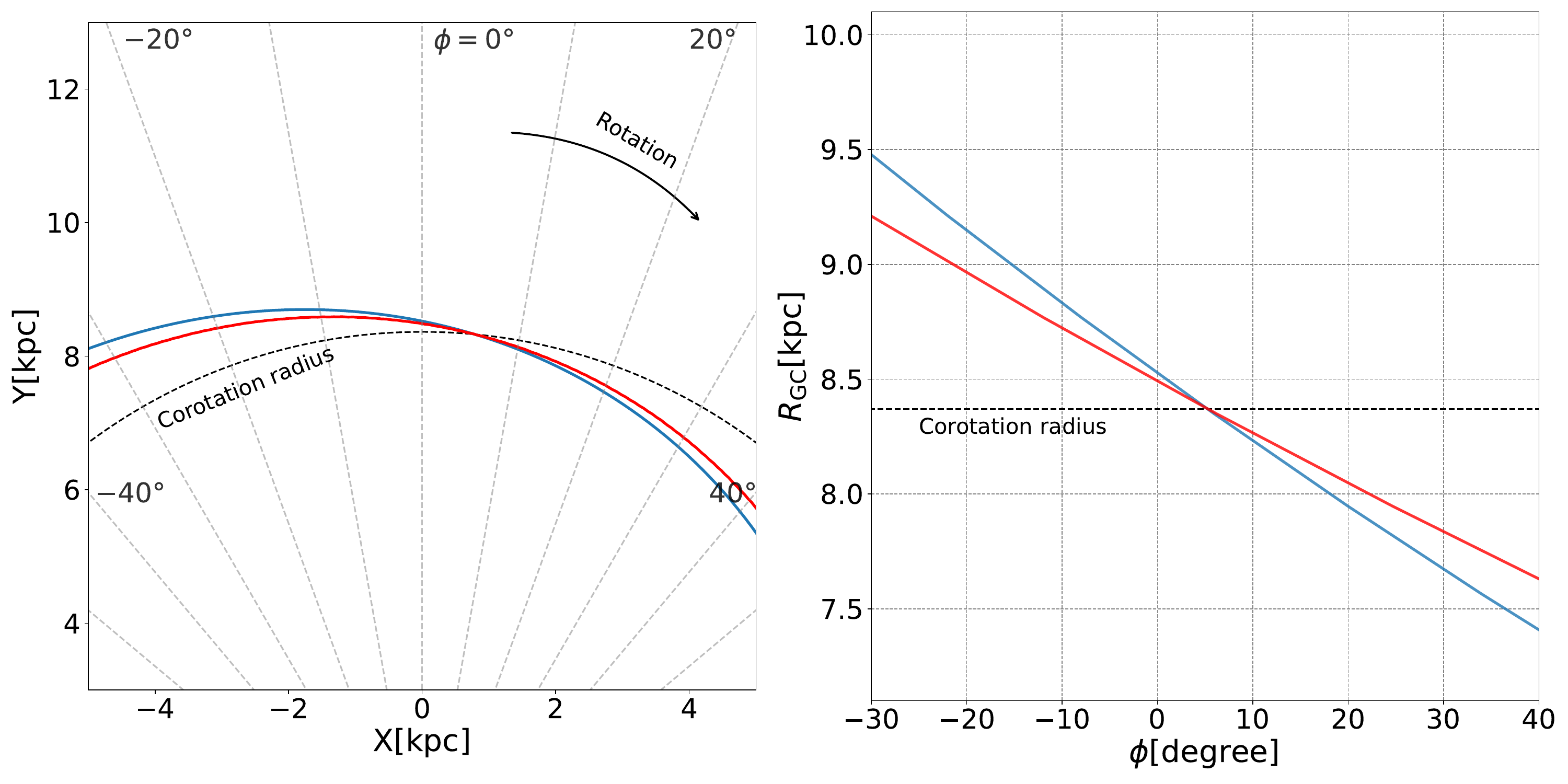}
    \caption{Sketch map of an expected age pattern of spiral arms under quasi-stationary scenario, depicting initial configurations (blue) against those 50 Myr later (red). Rotation curve parameters are sourced from~\citet{Reid19}, whereas the applied pattern speed, 28.2 km s$^{-1}$~kpc$^{-1}$, is as determined by~\citet{Dias19}.
}
    \label{fig_2}
\end{center}
\end{figure*}

In the age pattern schematic (Figure~\ref{fig_2}), according to~\citet{Reid19}, we set the sun's distance from the Galactic center at 8.15~kpc and its circular rotation speed at 236 km s$^{-1}$. These parameters were input into the MWPotential2014 model in $Galpy$~\citep{Bovy2013,Bovy2015} to obtain the rotation curve of the Milky Way. Subsequently, we calculated the circular rotation speed for each radius based on the rotation curve and subtracted a fixed spiral arm pattern speed of 28.2 km s$^{-1}$~kpc$^{-1}$, as established by~\citet{Dias19}. This result was then multiplied by 50~Myr to determine the predicted offset of the old stellar arm relative to the spiral arm according to the quasi-stationary density wave theory.

\bsp	
\label{lastpage}
\end{document}